\documentclass[12pt]{article}
\usepackage{latexsym}
\usepackage{epsfig,amssymb,euscript,slashed}
\usepackage[linktocpage]{hyperref}
\usepackage{amsmath}
\usepackage{cite} 
\usepackage{array,calc,epsfig}
\usepackage{bbm}
\usepackage{fancybox}

\numberwithin{equation}{section} %%
\usepackage[]{graphicx}
\usepackage{color}

\begin{document}
\font\cmss=cmss10 \font\cmsss=cmss10 at 7pt

\begin{flushright}{
%\scriptsize DFPD-17-TH-xx \\  
\scriptsize QMUL-PH-19-13}
\end{flushright}
\hfill
\vspace{14pt}
\begin{center}
{\Large 
\textbf{Holographic correlators in AdS$_3$\\ without Witten diagrams
}}

\end{center}

\vspace{8pt}
\begin{center}
{\textsl{Stefano Giusto$^{\,1, 2}$, Rodolfo Russo$^{\,3}$, Alexander Tyukov$^{\,1, 2}$ and Congkao Wen$^{\,3}$}}

\vspace{1cm}

\textit{\small ${}^{1}$ Dipartimento di Fisica ed Astronomia ``Galileo Galilei",  Universit\`a di Padova,\\Via Marzolo 8, 35131 Padova, Italy} \\  \vspace{6pt}

\textit{\small ${}^{2}$ I.N.F.N. Sezione di Padova,
Via Marzolo 8, 35131 Padova, Italy}\\
\vspace{6pt}

\textit{\small ${}^{3}$ Centre for Research in String Theory, School of Physics and Astronomy\\
Queen Mary University of London,
Mile End Road, London, E1 4NS,
United Kingdom}\\
\vspace{6pt}

\end{center}

\vspace{12pt}

\begin{center}
\textbf{Abstract}
\end{center}

\vspace{4pt} {\small
  \noindent
We present a formula for the holographic 4-point correlators in AdS$_3 \times S^3$ involving four single-trace operators of dimension $k, k, l, l$. As an input we use the supergravity results for the Heavy-Heavy-Light-Light correlators that can be derived by studying the linear fluctuations around known asymptotically AdS$_3 \times S^3$ geometries. When the operators of dimension $k$ and $l$ are in the same multiplet there are contributions due to the exchange of single-trace operators in the $t$ and $u$-channels, which are not captured by the approach mentioned above. However by rewriting the $s$-channel results in Mellin space we obtain a compact expression for the $s$-channel contribution that makes it possible to conjecture a formula for the complete result. We discuss some consistency checks that our proposal meets.

\vspace{1cm}

\thispagestyle{empty}

\vfill
\vskip 5.mm
\hrule width 5.cm
\vskip 2.mm
{
\noindent  {\scriptsize e-mails:  {\tt stefano.giusto@pd.infn.it, r.russo@qmul.ac.uk, tyukov@pd.infn.it, c.wen@qmul.ac.uk} }
}

\setcounter{footnote}{0}
\setcounter{page}{0}

\newpage

%%%%%%%%%%%%%%%%%%%%%%%%%%%%%%%%%%%%%%%%%

\section{Introduction}

Four point correlators of primary operators are key observables that package in a neat way the dynamical data of a Conformal Field Theory (CFT). In holographic theories, there is by definition a regime where the CFT can be described by a weakly coupled gravitational theory in AdS that captures a sector of primary operators corresponding to single particle excitations in the gravitational description. As common in the literature we refer to these operators as ``single-trace'', in analogy with the ${\cal N}=4$ SYM case, and as ``light'', since their conformal dimension remains finite in the large central charge limit. Four point correlators of single-trace Chiral Primary Operators (CPOs) in the gravity regime have been thoroughly studied in supersymmetric holographic CFTs, see for instance~\cite{DHoker:2002nbb} (and references therein) for a review of the early analysis of the ${\cal N}=4$ SYM case. In the last few years this problem has been readdressed by using new approaches exploiting a variety of techniques, such as the R-symmetry Ward Identities~\cite{Eden:2000bk}, the Mellin space formulation~\cite{Mack:2009mi,Penedones:2010ue}, the large spin perturbation theory~\cite{Alday:2016njk,Alday:2017vkk} and the Lorentzian inversion formula~\cite{Caron-Huot:2017vep}\!,\!\!\cite{Alday:2017vkk}. Again most results are obtained in the context of ${\cal N}=4$ SYM, where it is possible to write compact formulas for general four point correlators in the leading supergravity approximation~\cite{Rastelli:2016nze,Rastelli:2017udc,Aprile:2018efk,Caron-Huot:2018kta}, but these new approaches have been extended also to AdS$_7$/CFT$_6$~\cite{Zhou:2017zaw, Rastelli:2017ymc, Chester:2018dga, Abl:2019jhh} and AdS$_4$/CFT$_3$~\cite{Zhou:2017zaw, Chester:2018aca, Chester:2018lbz, Binder:2018yvd}.

In comparison, not much is known about supergravity correlators in other cases, and in this paper we focus on the prototypical AdS$_3$/CFT$_2$ duality already discussed in the original work by Maldacena~\cite{Maldacena:1997re}. We will thus consider the CFT that is holographically dual to type IIB string theory on AdS$_3\times S^3\times \mathcal{M}$, with $\mathcal{M}$ either $T^4$ or $K3$, commonly known as the D1-D5 CFT. The CFT spectrum contains CPOs of left/right dimension $(k/2,k/2)$ for any positive integer $k$. These operators are the analogue of the one-half BPS single-trace operators of $\mathcal{N}=4$ super Yang-Mills (SYM) theory, with one notable difference: there are actually $h^{1,1}(\mathcal{M})+1$ CPOs, $O^{(k,f)}$ (with $f=1,\ldots,h^{1,1}(\mathcal{M})+1$), for any integer $k\ge 1$, where $h^{p,q}$ are the Hodge numbers of $\mathcal{M}$, and an extra CPO, $\tilde O^{(k)}$, for any $k\ge 2$. A first example of AdS$_3$ correlators among four CPOs with $k=1$ was obtained in~\cite{Giusto:2018ovt}. The main result of the present work is to obtain explicit expressions for the correlators $\langle O^{(k,f)}_1 O^{(k,f)}_2 O^{(l,g)}_3 O^{(l,g)}_4 \rangle$, where $k,\,l=1,2,\ldots$ specify the dimension of the operators involved and $f,\,g=1,\ldots h^{1,1}(\mathcal{M})+1$ specify the flavour of the operators. We will consider two classes of correlators: those where all operators have the same flavour and those where the flavours $f$ and $g$ are different. In the second case only the $s$-channel decomposition (where $z_3\to z_4$) includes contributions from single-trace CPOs, while in the first case also the $t$ ($z_2\to z_3$) and the $u$ ($z_2 \to z_4$) channels receive non-trivial contributions from some single-trace CPOs. 

The approach we take in this paper generalises the technique developed in~\cite{Giusto:2018ovt}, where the correlators of CPOs with $k=1$ were derived by studying the supergravity quadratic fluctuation around a non-trivial geometry whose asymptotic limit is AdS$_3\times S^3\times \mathcal{M}$. The idea is to first consider a four point correlator involving two single-trace CPO, $O^{(l,g)}_3$ and $O^{(l,g)}_4$, and two multi-trace operators that are obtained by taking $N_0$ copies of $O^{(k,f)}_1$ at $z_1$ and $N_0$ copies of $O^{(k,f)}_2$ at $z_2$. When $N_0$ is of the order of the central charge $c=6N$ of the theory, these multi-trace operators become ``heavy'' and can be described by a classical regular solution~\cite{Lunin:2001jy,Kanitscheider:2007wq}. The heavy operators we use are obtained by superimposing the particular CPO associated with the K\"ahler form, which yields a particularly manageable supergravity solution (see for instance~(3.8) of~\cite{Bombini:2017sge}), however we believe our final results apply to any of the $h^{1,1}(\mathcal{M})+1$ CPOs $O^{(k,f)}$. This approach bypasses the need of the non-linear analysis around AdS$_3$ and made it possible to calculate explicitly several Heavy-Heavy-Light-Light (HHLL) correlators~\cite{Galliani:2016cai,Galliani:2017jlg,Bombini:2019vnc,Tian:2019ash,Bena:2019azk}. In order to obtain a standard correlator with single-trace CPOs we follow the logic of~\cite{Giusto:2018ovt} and take the $N_0/c\ll 1$ limit, assuming that in this regime the HHLL correlators obtained with the geometric approach mentioned above capture the contributions of all single-trace operators exchanged in the $s$-channel. This assumption is corroborated by the OPE checks we have done so far. If this approach is applied to a case where the flavours $f$ and $g$ are different, the result obtained is the full Light-Light-Light-Light (LLLL) correlator with single-trace CPOs. If instead we are considering a correlator with four CPOs of the same flavour, the light limit of the HHLL result misses the single-trace contributions in the $t$ and $u$-channels. We add these contributions starting from the partial ($s$-channel) result by using symmetries and constraints from the OPE expansion, as well as the structure of the correlators in Mellin space.

In section~\ref{sec:HHLL} we discuss how to derive the contributions to the $s$-channel for the correlators with four single-trace operators $\langle O^{(k,f)}_1 O^{(k,f)}_2 O^{(l,g)}_3 O^{(l,g)}_4 \rangle$ from the HHLL correlators computed in supergravity. In section~\ref{sec:corr-mellin} we obtain the Mellin transform of the previous results showing that they take a particularly simple form. When the flavours $f$ and $g$ are identical we present a conjecture for the full correlator including the $t$ and $u$-channel contributions and discuss checks that support our proposal.

\vspace{.2cm}
{\bf Note:} While we were finalising this paper we learned about~\cite{Rastelli:2019gtj} where longstanding difficulties with the exchange Witten diagrams in AdS$_3$ are resolved and AdS$_3 \times S^3$ holographic correlators are calculated utilizing a possible hidden symmetry.

\section{HHLL Correlators from supergravity}
\label{sec:HHLL}

Consider a CPO of the D1D5 CFT belonging to an $SU(2)_L\times SU(2)_R$ R-symmetry multiplet with spin $(k/2,k/2)$; we will denote a generic element of this multiplet as
\begin{equation}
O_k^{\alpha_1\ldots \alpha_k, \dot{\alpha}_1\ldots \dot{\alpha}_k} (z, \bar{z})\,,
\end{equation}
where $\alpha_i = \pm$ ($\dot\alpha_i = \pm$) are indices in the fundamental representation of $SU(2)_L$ ($SU(2)_R$) and $O_k^{\alpha_1\ldots \alpha_k, \dot{\alpha}_1\ldots \dot{\alpha}_k}$ is symmetric in the indices $\alpha_i$ and $\dot\alpha_i$. Adapting a formalism that is commonly used for $\mathcal{N}=4$ SYM, one can deal with the full R-symmetry multiplet by introducing the two-dimensional vectors $A^i_\alpha$ and $\bar A^i_{\dot \alpha}$, with the index $i$ going over the operators appearing in the correlator, and define
\begin{equation}
  \label{eq:Aalpha}
O_i^{(k)} \equiv A^{i}_{\alpha_1}\ldots A^{i}_{\alpha_k} \bar{A}^{i}_{\dot{\alpha}_1}\ldots \bar{A}^{i}_{\dot{\alpha}_k} O_k^{\alpha_1\ldots \alpha_k, \dot{\alpha}_1\ldots \dot{\alpha}_k} (z_i, \bar{z}_i).
\end{equation}
Global conformal and R-symmetry invariance implies that a general four-point amplitude containing operators of dimensions $k$ and $l$ has the form
\begin{equation}
\label{eq:LLLLcor}
\mathcal{C}_{k,l}(\alpha_c,\bar \alpha_c,z,\bar z)\equiv \langle O^{(k)}_1 O^{(k)}_2 O^{(l)}_3 O^{(l)}_4 \rangle = \left(\frac{|A^1\cdot A^2|^2}{|z_{12}|^2}\right)^{k}\left(\frac{|A^3\cdot A^4|^2}{|z_{34}|^2}\right)^{l} \, \mathcal{G}_{k,l}(\alpha_c, {\bar \alpha_c}, z, \bar{z})\,,
\end{equation}
with
\begin{equation}\label{eq:ratios}
z_{ij}\equiv z_i - z_j\,,\quad A^i \cdot A^j \equiv \epsilon_{\alpha\beta} A^i_\alpha A^j_\beta\,;
\end{equation}
$z$ and $\alpha_c$ are the cross-ratios
\begin{equation} \label{eq:sitau}
  \begin{aligned}
    z\equiv & \frac{z_{14} z_{23}}{z_{13} z_{24}}\,,\quad V\equiv |z|^2\,, \quad U\equiv |1-z|^2\,,
\\
    \alpha_c\equiv & \frac{A^1\cdot A^3\,A^2\cdot A^4}{A^1\cdot A^4\,A^2\cdot A^3} \,,\quad    \sigma\equiv \left| {\alpha_c  \over 1- \alpha_c} \right|^2, \quad  \tau\equiv \left| { 1 \over 1-\alpha_c} \right|^2  \,,
  \end{aligned}
  \end{equation}
where for later convenience we defined also the variables $U$, $V$, $\sigma$, and $\tau$. Note that we can reconstruct the full $\alpha_c$ dependence of the correlator by taking
\begin{equation}\label{eq:choiceA}
A^1 = \bar{A}^1= \begin{pmatrix} 0 \\1\end{pmatrix}\!,\, A^2 = \bar{A}^2= \begin{pmatrix} 1 \\0\end{pmatrix}\!,\, A^3 =  \begin{pmatrix} 1 \\ \frac{-i}{\sqrt{\alpha_c}}\end{pmatrix}\!,\,\bar{A}^3 =  \begin{pmatrix} 1 \\ \frac{-i}{\sqrt{\bar \alpha_c}}\end{pmatrix}\!, \,A^4 =  \begin{pmatrix}  \frac{i}{\sqrt{\alpha_c}} \\ 1\end{pmatrix}\!,\,\bar{A}^4 =  \begin{pmatrix} \frac{i}{\sqrt{\bar \alpha_c}}\\1 \end{pmatrix}\!.
\end{equation}
This choice agrees with the definition of $\alpha_c$ in \eqref{eq:sitau} and also guarantees that $O^{(k)}_1=(O^{(k)}_2)^\dagger$ and $O^{(l)}_3=(O^{(l)}_4)^\dagger$. Similarly one can choose $z_1=0$, $z_2\to \infty$, $z_3=1$. Then we have $z_4=z$, $|z_{34}|^2=|1-z|^2$, $|A^1 \cdot A^2|^2=1$ and  $|A^3 \cdot A^4|^2=|\frac{1-\alpha_c}{\alpha_c}|^2$.

Our goal is to compute the function $\mathcal{G}_{k,l}(\alpha_c, \bar{\alpha}_c, z, \bar{z})$ at the strong-coupling point of the CFT dual to classical supergravity. In this limit the operators $O^{(l)}$ are dual to supergravity fields describing small perturbations around the vacuum, represented by the global AdS$_3\times S^3\times \mathcal{M}$ geometry, and satisfying a linear wave equation. The operators $O^{(k)}$ inserted at $z_1=0$ and $z_2=\infty$ can be thought to create and destroy a certain light CFT state of dimension $k\ll N$. However, as mentioned in the introduction, we deform these insertions by taking $N_0$ copies of each one of the two operators $O^{(k)}_{1,2}$, with $N_0$ a large number of order $N$, and then we flow to the Ramond (R) sector. The final state, which we can denote by $|k,N_0\rangle_R$, is a Ramond ground state with conformal dimension $(N/4,N/4)$, which admits a gravitational description in terms of a non-trivial solution of classical supergravity whose asymptotic limit is AdS$_3\times S^3\times \mathcal{M}$ (see for instance~(3.8) of~\cite{Bombini:2017sge}). By solving the wave equation associated with the operator $O^{(l)}$ in the geometry dual to the ``heavy" state $|k,N_0\rangle_R$, one can compute the HHLL correlator \cite{Galliani:2017jlg}
\begin{equation}\label{eq:HHLLcor}
\mathcal{C}_{k,l}^R \equiv {}_R\langle k,N_0| O^{(l)}_3 O^{(l)}_4 |k,N_0\rangle_R\,.
\end{equation}
The LLLL correlator defined in \eqref{eq:LLLLcor} should be linked with the $N_0\to 1$ limit of the HHLL correlator \eqref{eq:HHLLcor}, after flowing back to the Neveu-Schwarz (NS) sector:
\begin{equation}\label{eq:HHLL2LLLL}
{}_R\langle k,N_0| O^{(l)}_3 O^{(l)}_4 |k,N_0\rangle_R \stackrel{\mathrm{spectral~flow}}{\longrightarrow} {}_{NS}\langle k,N_0| O^{(l)}_3 O^{(l)}_4 |k,N_0\rangle_{NS}\stackrel{N_0\to 1}{\longrightarrow} \langle O^{(k)}_1 O^{(k)}_2 O^{(l)}_3 O^{(l)}_4 \rangle \,.
\end{equation}
As explained in \cite{Giusto:2018ovt}, the procedure above does not produce the full LLLL correlator, but only its $s$-channel contribution, i.e. the terms of the correlator associated with the exchange of all single-trace operators in the limit $z_3\to z_4$ (or $z\to 1$). In this section we concentrate on extracting this contribution from the gravity result for the HHLL correlator \eqref{eq:HHLLcor}. The problem of reconstructing the full LLLL correlator from the $s$-channel contribution will be discussed in the next section.

The direct gravity derivation of the HHLL correlator \eqref{eq:HHLLcor} is hampered by the fact that the wave equation for the operator $O^{(l)}$ is not simple. A simpler wave equation is obtained by considering the supersymmetry descendants of the CPOs, obtained by acting with the left and right supercharges $G^\alpha_A$, $\tilde G^{\dot \alpha}_A$: here $A=1,2$ is an index of the ``bonus" $SU(2)$ symmetry of the CFT; it will not play any role in our analysis and we can just take $A=1$ in the following. The descendant of the operator $O^{(l)}_i$ is an operator of dimension $(l/2+1/2,l/2+1/2)$ and spin $(l/2-1/2,l/2-1/2)$ which we will denote by
\begin{equation}
B_i^{(l-1)} \equiv A^{i}_{\alpha_1}\ldots A^{i}_{\alpha_{l-1}} \bar{A}^{i}_{\dot{\alpha}_1}\ldots \bar{A}^{i}_{\dot{\alpha}_{l-1}} 
B_{l-1}^{\alpha_1\ldots \alpha_{l-1}, \dot{\alpha}_1\ldots \dot{\alpha}_{l-1}} (z_i, \bar{z}_i)\,.
\end{equation}
One can compactly write the relation between the full R-symmetry multiplets $O^{(l)}_i$ and $B^{(l-1)}_i$ by defining
\begin{equation}
G\equiv G^+_{1,-1/2} - \frac{i}{\sqrt{\alpha_c}}\, G^-_{1,-1/2} \,,\quad \tilde G\equiv \tilde G^+_{1,-1/2}  - \frac{i}{\sqrt{\bar \alpha_c}}\, \tilde G^-_{1,-1/2}\,,
\end{equation}
where the index $-1/2$ denotes the mode of the current. Then one has
\begin{equation}\label{eq:BO}
G \tilde G B_3^{(l-1)} = O_3^{(l)} \,,\quad G \tilde G O_4^{(l)} = \frac{|1-\alpha_c|^2}{|\alpha_c|^2} B_4^{(l-1)}\,,
\end{equation}
where we have made explicit use of the choice \eqref{eq:choiceA}. 
The operators $B^{(l-1)}$ are dual to minimally coupled scalars in the 6D theory reduced on $\mathcal{M}$. By solving the corresponding wave equation in the geometry dual to the heavy state $|k,N_0\rangle_R$ one can derive the correlator
\begin{equation}
\widehat{\mathcal{C}}^{\,R}_{k,l}\equiv {}_R\langle k,N_0| B^{(l)}_3 B^{(l)}_4 |k,N_0\rangle_R\,.
\end{equation}
Using the relation between $O^{(l)}$ and $B^{(l-1)}$ in \eqref{eq:BO}, and the fact that the R ground state $|k,N_0\rangle_R$ is annihilated by both the supercharges $G$ and $\tilde G$, it is immediate to derive \cite{Bombini:2017sge} a Ward identity relating the HHLL correlators $\mathcal{C}_{k,l}^R$ and $\widehat{\mathcal{C}}^{\,R}_{k,l-1}$:
\begin{equation}\label{eq:WI}
\frac{|1-\alpha_c|^2}{|\alpha_c|^2} \,|z|^{l-1} \,\widehat{\mathcal{C}}^{\,R}_{k,l-1} = \partial {\bar \partial}\, (|z|^l\, \mathcal{C}_{k,l}^R)\,.
\end{equation}
Analogously to what is explained in \eqref{eq:HHLL2LLLL}, the HHLL correlator $\widehat{\mathcal{C}}^{\,R}_{k,l}$ can be related with a LLLL 4-point function $\widehat{\mathcal{C}}_{k,l}$ between two CPOs and two descendants
\begin{equation}
\widehat{\mathcal{C}}_{k,l}(\alpha_c,\bar \alpha_c,z,\bar z)\equiv \langle O^{(k)}_1 O^{(k)}_2 B^{(l)}_3 B^{(l)}_4 \rangle\,,
\end{equation}
by performing the spectral flow\footnote{Under spectral flow an operator of spin $(m,\bar m)$ transforms as $O^{m\bar m}_R(z,\bar z)= z^{-m} {\bar z}^{-\bar m} \,O^{m\bar m}_{NS}(z,\bar z)$ and thus the transformation from R to NS sector is implemented by multiplying the $\alpha_c$-dependent operator $O^{(l)}_4$ by the factor $|z|^{-l}$ and by sending $(\alpha_c,\bar \alpha_c)\to  (\alpha_c z^{-1},\bar \alpha_c \bar z^{-1})$ (and analogously for $B^{(l)}_4$).} and taking the $N_0\to 1$ limit:
\begin{equation}\label{eq:N0toone}
\widehat{\mathcal{C}}_{k,l}(\alpha_c \,z^{-1},\bar \alpha_c \,{\bar z}^{-1},z,\bar z) = \lim_{N_0\to 1} |z|^l\,\widehat{\mathcal{C}}^{\,R}_{k,l}\,.
\end{equation}
The Ward identity \eqref{eq:WI} can then be translated into an identity for the LLLL correlators
\begin{equation}\label{eq:WIbis}
\frac{|1-\alpha_c|^2}{|\alpha_c|^2} \,\widehat{\mathcal{C}}_{k,l-1}(\alpha_c \,z^{-1},\bar \alpha_c \,{\bar z}^{-1},z,\bar z) =\partial {\bar \partial} \,\mathcal{C}_{k,l}(\alpha_c \,z^{-1},\bar \alpha_c \,{\bar z}^{-1},z,\bar z) \,.
\end{equation}
Summarising, the gravity computation provides the HHLL correlator $\widehat{\mathcal{C}}^{\,R}_{k,l-1}$, from which one deduces the LLLL correlator $\widehat{\mathcal{C}}_{k,l-1}$ via \eqref{eq:N0toone} and finally the $s$-channel contribution to the correlator of four CPOs $\mathcal{C}_{k,l}$ by solving the Ward identity \eqref{eq:WIbis}.

We first illustrate this procedure in the simplest case $l=1$ and then generalize to $l\ge 1$. Since the operator $B_0$ has no spin, the gravity input $\widehat{\mathcal{C}}^{\,R}_{k,0}$ does not depend on $\alpha_c$: the $N_0\to 1$ limit of this HHLL correlator was computed already in \cite{Bombini:2017sge} and the result can be written in the form 
\begin{equation}\label{eq:gravl0}
\widehat{\mathcal{C}}_{k,0}(z,\bar z)=\partial \bar \partial \left[ \left(1-\frac{k}{N}\right)|1-z|^{-2}+\frac{2k}{N \,\pi} |z|^2 \sum_{p=1}^k \frac{1}{p} \hat D_{p,p,2,2}\right]\,,
\end{equation}
where the $\hat D$-functions are defined in \eqref{eq:hatDM}}; see appendix~A of~\cite{Giusto:2018ovt} for more details on our conventions. Then the Ward identity \eqref{eq:WIbis} is solved\footnote{The solution of the differential equation \eqref{eq:WIbis} for $\mathcal{C}_{k,l}$ is of course non-unique: the ambiguity has been used to choose the term $\mathcal{C}_{k,1}^{(0)}$ to be consistent with the $z\to 1$ OPE of the free CFT.} by
\begin{equation}
  \label{eq:Ck1}
\mathcal{C}_{k,1}(\alpha_c ,\bar \alpha_c ,z,\bar z) =\mathcal{C}_{k,1}^{(0)}+\frac{2k}{N \,\pi} \frac{|1-\alpha_c\,z|^2}{|\alpha_c|^2} \sum_{p=1}^k \frac{1}{p} \hat D_{p,p,2,2}\,,
\end{equation}
where we have isolated the term $\mathcal{C}_{k,1}^{(0)}$, which does not contain $\hat D$-functions:
\begin{equation}
\mathcal{C}_{k,1}^{(0)}\equiv \left(1-\frac{k}{N}\right) \frac{|1-\alpha_c|^2}{|\alpha_c|^2} \frac{1}{|1-z|^2} \,.
\end{equation}
The $\alpha_c$-dependence of the $\hat D$-dependent part of the  correlator is particular simple for $l=1$, as it reduces to an overall factor $\frac{|1-\alpha_c\,z|^2}{|1-\alpha_c|^2}$, which descends from the ``kinematical'' factor $\frac{|1-\alpha_c|^2}{|\alpha_c|^2} $ on the l.h.s. of \eqref{eq:WIbis}. This overall kinematical factor will always be present even for general $l$, so it is convenient to use the definition \eqref{eq:LLLLcor} to obtain ${\cal G}_{k,l}$ and then define a stripped amplitude as follows
\begin{equation}
  \label{eq:theform}
  \mathcal{G}_{k,l}(\alpha_c ,\bar \alpha_c ,z,\bar z)
  \equiv \mathcal{G}^{(0)}_{k,l} + \frac{1}{N} \frac{|1-\alpha_c\,z|^2}{|1-\alpha_c|^2} \widetilde{\mathcal{G}}_{k,l}\,, 
\end{equation}
where $\mathcal{G}^{(0)}_{k,l}$ is a rational contribution and $\widetilde{\mathcal{G}}_{k,l}$ is the stripped amplitude that contains the non-trivial term dependent on the $\hat D$-function. Since by definition correlators are polynomial function of $1/\alpha_c$, $\widetilde{\mathcal{G}}_{k,l}$ is regular as $\alpha_c\to 1/z$. Hence, as already noted in \cite{Giusto:2018ovt} in the AdS$_3$ context, the form~\eqref{eq:theform} guarantees that $\mathcal{G}$ satisfies an analogue of the $\mathcal{N}=4$ superconformal Ward identity:
\begin{equation}\label{eq:superWI}
\partial_{\bar z} \left (\mathcal{G}_{k,l}(\alpha_c ,\bar \alpha_c ,z,\bar z)|_{\bar \alpha_c\to 1/\bar z} \right)=0\,.
\end{equation}

For $l=1$ we can use~\eqref{eq:Ck1} and obtain
\begin{equation} \label{eq:stak1}
\mathcal{G}^{(0,s)}_{k,1}=\left(1-\frac{k}{N}\right) \,, \quad \widetilde{\mathcal{G}}^{(s)}_{k,1}(z,\bar z) = \frac{2k}{\pi} |1-z|^2\,\sum_{p=1}^k \frac{1}{p} \hat D_{p,p,2,2} \,.
\end{equation}
Here the superscript $(s)$ emphasizes that this is only the $s$-channel contribution to the full correlator.

%%%%%%%%%%

The $\alpha_c$-dependence for $l>1$ is more involved because the gravity input $\widehat{\mathcal{C}}^{\,R}_{k,l-1}$ itself depends on $\alpha_c$. The supergravity computation of $\widehat{\mathcal{C}}^{\,R}_{k,l-1}$ for $N_0\to 1$ is a generalization of the one performed in \cite{Bombini:2017sge}. The geometry dual to the R ground state $|k,N_0\rangle_R$ is  given for example in eqs. (2.1) and (3.11) of \cite{Bena:2015bea}: the supergravity parameters $a$ and $b$ are related to the CFT integers $N_0$, $N$ and $k$ by $b^2/(a^2+\frac{b^2}{2}) = k N_0/N$. As we are interested in the limit with $k$ fixed and $N_0/N\ll 1$, one should take the limit $b/a\ll 1$, in which the geometry is a small deformation of global AdS$_3\times S^3$. The CPO descendant $B^{(l-1)}$ is dual to a minimally coupled scalar $\Phi$ in 6D, which satisfies the equation $\Box \Phi=0$, where the d'Alambertian operator is constructed with the 6D Einstein metric; both $\Phi$ and $\Box$ can be expanded in $b/a$
\begin{equation}
\Phi=\Phi_0 + b^2 \Phi_1\,,\quad \Box=\Box_0 +b^2 \Box_1\,.
\end{equation}
The zero-th order solution $\Phi_0$ is a harmonic function in AdS$_3\times S^3$, which we can take of the form
\begin{equation}
\Phi_0 = B_0 \,Y_{l-1}^{\alpha_c,\bar \alpha_c}\,,
\end{equation}
where $B_0$ is an eigenfunction of the AdS$_3$ Laplacian $\Box_{AdS_3} B_0 =(l-1)(l+1)B_0$ and $Y_{l-1}^{\alpha_c,\bar \alpha_c}$ is an $\alpha_c$-dependent superposition of spherical harmonics of spin $l-1$:
\begin{equation}\label{eq:sphericalharm}
Y_{l-1}^{\alpha_c,\bar \alpha_c}\equiv \frac{\sqrt{l}}{\sqrt{2}\, \pi}(A^3_\alpha \bar A^3_{\dot \alpha} Y^{\alpha\dot\alpha})^{l-1}\,;
\end{equation}
here the vectors $A^3$ and $\bar A^3$ are the ones defined in \eqref{eq:choiceA}, and $Y^{\alpha\dot\alpha}$ are spherical harmonics of spin $(1/2,1/2)$; for example, in the standard Cartan parametrization of $S^3$, one can take
\begin{equation}
Y^{++} = \sin\theta e^{i\hat \phi}\,,\quad Y^{+-} = \cos\theta e^{i\hat \psi}\,,\quad Y^{-+} =- \cos\theta e^{-i\hat \psi}\,,\quad Y^{--} = \sin\theta e^{-i\hat \phi}\,.
\end{equation}
With these choices, the spherical harmonics defined by \eqref{eq:sphericalharm} have norm $1$. The first order solution $\Phi_1$ is found by solving
\begin{equation}\label{eq:wave1}
\Box_0 \Phi_1 = - \Box_1 \Phi_0\,.
\end{equation}
As the wave equation with $b\not=0$ is non-separable (for generic $k$), $\Phi_1$ is a superposition of spherical harmonics of different spin; to extract the correlator where the two light operators are the same, one should project $\Phi_1$ onto $Y_{l-1}$: the projected wave function $\Phi_1^{(l-1)}$ satisfies the equation
\begin{equation}
\Box_{AdS_3} \Phi_1^{(l-1)} = -\int d\Omega_3\, (Y_{l-1}^{\alpha_c,\bar \alpha_c})^* \Box_1 \Phi_0\,,
\end{equation}
which can be easily solved using the AdS$_3$ bulk-to-bulk propagator. The HHLL correlator $\widehat{\mathcal{C}}^{\,R}_{k,l-1}$ is extracted from normalizable term of $\Phi_1^{(l-1)}$ in the expansion around the AdS boundary:
\begin{equation}
\Phi_1^{(l-1)}(r,z,\bar z) \stackrel{r\to\infty}{\longrightarrow} \frac{|z|^{l+1} \,\widehat{\mathcal{C}}^{\,R}_{k,l-1}(\alpha_c,\bar\alpha_c,z,\bar z)}{r^{l+1}}\,,
\end{equation}
where $r$ and $z,\bar z$ are the radial and boundary coordinates of AdS$_3$. Further details of the gravity computation will be given in a future work \cite{GRTW}. After using \eqref{eq:N0toone}, the result of the computation for generic $l$ can be written in a form that is structurally similar to \eqref{eq:gravl0}:
\begin{equation}\label{eq:Chatkl}
\begin{aligned}
\widehat{\mathcal{C}}_{k,l}(\alpha_c z^{-1},\bar \alpha_c \bar z^{-1}, z,\bar z)=&\left(1-\frac{k\,l}{N}\right) \frac{|z-\alpha_c|^{2(l-1)}}{|\alpha_c|^{2(l-1)}} \partial \bar \partial \left(\frac{1}{|1-z|^{2 l}}\right)\\
&+\partial \bar \partial \left[\frac{2k l^2}{N \,\pi} |z|^2 \sum_{p=1}^k c_{p,l}(\alpha_c z^{-1},\bar \alpha_c \bar z^{-1}) \hat D_{p,p,l+1,l+1}\right]\,,
\end{aligned}
\end{equation}
where the $\alpha_c$-dependent coefficients $c_{p,l}(\alpha_c,\bar \alpha_c)$ are defined by
\begin{equation}\label{eq:defcp}
c_{p,l}(\alpha_c,\bar \alpha_c) = \int d\Omega_3 \, |Y_{l-1}^{\alpha_c,\bar \alpha_c}|^2 \sin^{2(p-1)}\theta\,.
\end{equation}
These $c_{p,l}$'s take a particular compact form when written in terms of the variables $\sigma$,$ \tau$ defined in~\eqref{eq:sitau}:
\begin{equation}
  \label{eq:cst}
  c_{p,l}(\sigma,\tau) = \sum_{n=0}^{l-1}\sum_{m=0}^{n}\frac{\sigma^{n-m} \tau^m \binom{l-1}{n}\, \binom{p-1}{n}\, \binom{n}{m}^2} {\sigma^{l-1} \, \binom{p+l-1}{l} }\;.
\end{equation}
Using as before the Ward identity \eqref{eq:WIbis}, one deduces from \eqref{eq:Chatkl} the $s$-channel contribution for the 4-point function of CPOs of generic dimension $k$ and $l$:
\begin{equation}\label{eq:2.28}
\mathcal{G}^{(0,s)}_{k,l}=\left(1-\frac{k l}{N}\right)\,,\quad  \widetilde{\mathcal G}^{(s)}_{k,l}(\sigma ,\tau ,z,\bar z) =\frac{2k l^2}{\pi} \sigma^{l-1}\,|1-z|^{2l}\sum_{p=1}^k c_{p,l}(\sigma,\tau) \hat D_{p,p,l+1,l+1}\,.
\end{equation}
One can check that the coefficient of the leading $z,\bar{z}\to 1$ limit of $\mathcal{G}^{(s)}_{k,l}$ is normalised to $1$, reproducing the contribution of the identity, thanks to the cancellation of the $1/N$ terms between the rational contribution $\mathcal{G}^{(0,s)}_{k,l}$ and the rest. Notice that for $l>1$ the stripped amplitude $\mathcal{G}^{(s)}_{k,l}$ depends non-trivially on $\sigma$, $\tau$ both through the overall normalization $\sigma^{l-1}$ and the coefficients $c_{p,l}(\sigma,\tau)$ defined\footnote{One can check that $c_{p,1}(\sigma,\tau)=1/p$, so we recover the $\alpha$-independent result for $l=1$.} by \eqref{eq:cst}. Finally, even if it is not manifest in the formulation~\eqref{eq:2.28}, the result is symmetric in the exchange $k\leftrightarrow l$ as expected for this correlator. This will be evident in the Mellin formulation discussed in the next section.

\section{Correlators in Mellin space}
\label{sec:corr-mellin}

As in ${\cal N}=4$ SYM, it is convenient to rewrite the HHLL results obtained so far in Mellin space since this neatly selects the single-trace contributions we are interested in, as discussed in~\cite{Penedones:2010ue,Rastelli:2016nze,Rastelli:2017udc}. This can be done by using the Mellin representation of the $\hat{D}$-functions
  \begin{align} \label{eq:hatDM}
&\hat{D}_{\Delta_1, \Delta_2, \Delta_3,  \Delta_4}(z,\bar{z}) = \Gamma\left({\Delta_{1234} -d \over 2}\right){\pi^{d/2} \over 2 \prod_{j=1}^4 \Gamma(\Delta_j)} 
\int {ds \over 4\pi i} {dt \over 4\pi i} U^{s\over 2}  V^{t\over 2} \,\Gamma\left[-{s\over 2}\right] \Gamma\left[-{t\over 2}\right] \\
&\times \, \Gamma\left[\Delta_4 +{s+t \over 2}\right]\Gamma\left[{\Delta_{12}  -\Delta_{34} -s \over 2}\right]  \Gamma \left[{ \Delta_{23}  -\Delta_{14} -t \over 2} \right] \Gamma\left[{\Delta_{134} - \Delta_{2} +s + t \over 2}\right]\,, \nonumber
  \end{align}
  where $\Delta_{i_1\ldots i_n}=\Delta_{i_1} + \ldots + \Delta_{i_n}$. Then we can rewrite the stripped amplitudes derived in the previous section by using\footnote{As in the previous section we take $k\geq l$, but we will see that the final results do not depend on this assumption.}
  \begin{align}
    \label{eq:MTstramp}
    \widetilde{\cal G}_{k,l}= \frac{\pi}{2} \int \frac{ds}{4\pi i}\, \frac{dt}{4\pi i} U^{\frac{s}{2}} V^{\frac{t}{2}-\frac{k+l}{2}}\,\Gamma\left[k-\frac{s}{2}\right] \Gamma\left[l-\frac{s}{2}\right] \Gamma^2\left[\frac{k+l-t}{2}\right] \Gamma^2\left[\frac{k+l-\tilde{u}}{2}\right] \, \widetilde{\cal M}_{k, l}\;,
  \end{align}
  where $\tilde{u}=u-2=2k+2l-s-t-2$ and $ V=|z|^2$, $U=|1-z|^2$, and the amplitude $\widetilde{\cal M}_{k, l}$ is a function of the Mellin variables $s, t$. 

The $s$-channel contribution to the stripped amplitude~\eqref{eq:2.28} obtained from our supergravity approach is written as a linear combination of terms proportional to $|1-z|^{2l} \hat{D}_{p,p,l+1,l+1}$ with $p=1,\ldots,k$. Then from (\ref{eq:hatDM}) and (\ref{eq:MTstramp}), we obtain 
  \begin{equation} \label{eq:mtkk11}
 \frac{2}{\pi} |1-z|^{2l}   \hat{D}_{p,p,l+1,l+1}  \to  \frac{\Gamma(p+l)}{\Gamma^2(p)\Gamma^2(l+1)}  {\Gamma(p-1-s/2) \over \Gamma(k-s/2)} \, ,
  \end{equation}
where the arrow indicates the Mellin transform.  Combining the above result with the coefficient $c_{p, l}$ given in (\ref{eq:cst}), we find that the Mellin amplitude for $\widetilde{\mathcal G}^{(s)}_{k,l}$ in (\ref{eq:2.28}) takes a simple form\footnote{For later convenience we relabelled the sum over $n$ in~\eqref{eq:cst} in terms of $p=l-1-n$.}, 
  \begin{align} \label{eq:Ms-kl}
\widetilde{\cal M}^{(s)}_{k, l} (s, t; \sigma, \tau)= - {2 k l \over \Gamma(k)\Gamma(l)}  \sum_{p=0}^{l-1} { {l-1}\choose{l{-}1{-}p}}^2 \, 
\left( \prod^{l-1-p}_{i=1} { k{-} i \over l {-} i  } \right) \,  { Y({l{-}1{-}p}; \sigma, \tau) \over s - 2(l{-}1)+2p }\,,
\end{align}
where we define $\prod\limits^{0}_{i=1} \left(\ldots\right) \equiv 1$ and $Y(l{-}1{-}p; \sigma, \tau)$ is a degree-$(l{-}1{-}p)$ polynomial of $ \sigma, \tau$ 
\begin{align}
Y(n; \sigma, \tau)= \sum_{m = 0}^n { {n}\choose{m}} ^2 \sigma^{n-m} \tau^m \, .
\end{align}
The result \eqref{eq:Ms-kl} follows the standard pattern observed in other holographic correlators: the single-trace operators contribute to the supergravity correlator only if their twist is smaller than that of the first double-trace operator appearing in the OPE channel under consideration. 

If the operators of dimension $k$ have a different flavour than those of dimension $1$, then there are no new contributions from the exchange of single-trace operators in other channels and~\eqref{eq:Ms-kl} or~\eqref{eq:2.28} represent the full correlators. If on the other hand all operators have the same flavour, then it is necessary to add the contribution of the single-trace exchanges in the $t$ and $u$-channel. For the special case of $k=l$, the correlator is permutation symmetric, therefore knowing the result in $s$-channel is enough to fix the full correlator.  When $k=l$, $\widetilde{\cal M}^{(s)}_{k, l}$ reduces to
  \begin{align} \label{eq:Ms-kk}
\widetilde{\cal M}^{(s)}_{k, k} (s, t; \sigma, \tau)= - {2 k^2 \over \Gamma(k)^2 }  \sum_{p=0}^{k-1} { {k-1}\choose{k{-}1{-}p}}^2 \,  { Y({k{-}1{-}p}; \sigma, \tau) \over s - 2(k{-}1)+2p }\, .
\end{align}
The $t$-channel contribution may be obtained from $s$-channel result via exchange $1 \leftrightarrow 3$, from which we obtain
  \begin{align}
\widetilde{\cal M}^{(t)}_{k, k}(s, t; \sigma, \tau) =\tau^{k-1} \widetilde{\cal M}^{(s)}_{k, k} (t, s; \sigma/\tau, 1/\tau)\, .
\end{align}
Analogously, for the $u$-channel contribution, we have
  \begin{align}
\widetilde{\cal M}^{(u)}_{k, k}(s, t; \sigma, \tau) =\sigma^{k-1} \widetilde{\cal M}^{(s)}_{k, k} (\tilde{u}, t; 1/\sigma, \tau/\sigma)\, .
\end{align}
The full correlator in Mellin space is then the sum of all these contributions, 
  \begin{align} \label{eq:kk-stu}
\widetilde{\cal M}_{k, k}(s, t; \sigma, \tau) = \widetilde{\cal M}^{(s)}_{k, k} (s, t; \sigma, \tau) + \widetilde{\cal M}^{(t)}_{k, k} (s, t; \sigma, \tau) +\widetilde{\cal M}^{(u)}_{k, k} (s, t; \sigma, \tau)\, .
\end{align}
For the simplest case $k=1$, we have
  \begin{align} \label{eq:k=1-correlator}
\widetilde{\cal M}_{1, 1}(s, t; \sigma, \tau) =  -2 \left( {1 \over s} + {1 \over t} + {1 \over \tilde{u}} \right) \, ,
\end{align}
which reproduces the result obtained in~\cite{Giusto:2018ovt}. 

Non-trivially, $\widetilde{\cal M}_{k, k}(s, t; \sigma, \tau)$ can be recast into a more suggestive form, 
  \begin{align} \label{eq:kk-stu-2}
\widetilde{\cal M}_{k, k}(s, t; \sigma, \tau) = \sum_{p+q+r=k-1} b_{pqr} \,  \sigma^q \tau^r  \left(  { 1 \over s-s_M+2p} +{ 1 \over t - t_M + 2r } +{ 1 \over \tilde{u} - {u}_M+ 2q }  \right) \, ,
\end{align}
where $s_M = t_M  = u_M=2(k-1)$, and the coefficient $b_{pqr}$ is given by
 \begin{align} \label{eq:apqr-kk}
b_{pqr} = - {2 k^2 \over \Gamma(k)^2}  {k-1\choose p,q,r}^2\,, 
 \end{align}  
 where the parenthesis stands for the standard multinomial coefficient. The new expression for $\widetilde{\cal M}_{k, k}$
suggests an interesting structure for the Mellin amplitudes in ${\rm AdS}_3
\times S^3$, which will be important for the generalisation to
$\widetilde{\cal M}_{k, l}$, as we will discuss in the following
section. Finally, we remark that no ``contact terms" can be added to
\eqref{eq:kk-stu-2}, since in a two-derivative theory the Mellin
amplitude ${\cal M}$ (which is the Mellin transform of ${\mathcal G}$)
should be linear in the Mellin variables as $s,t,u \to \infty$. Then
one can deduce that the stripped amplitude $\widetilde{\cal M}$ should
go to zero as $1/s$. 
 
\subsection{General $\widetilde{\mathcal G}_{k,l}$ correlators}

For general $\widetilde{\mathcal G}_{k,l}$ correlators among operators of the same flavour, symmetries are not enough to fix the $t$- and $u$-channel contributions from $s$-channel results. As a first simple example we start from $l=1$. For this particular case, the stripped amplitude has no dependence on the R-symmetry variables $\sigma, \tau$, and is given by 
 \begin{align}
\widetilde{\cal M}^{(s)}_{k,1}=- { 2 k \over \Gamma(k)} {1 \over s} \, .
\end{align}

For $k=1$ the $t$ and $u$-channels are determined by the $s$-channel due to symmetry, as given in \eqref{eq:k=1-correlator}. The correlator in coordinate space was derived in~\cite{Giusto:2018ovt} and is given by
\begin{equation}\label{eq:11corr}
\mathcal{G}_{1,1} = \mathcal{G}^{(0)}_{1,1}+\frac{2 }{N\,\pi}\frac{|1-\alpha_c z|^2}{|1-\alpha_c|^2}|1-z|^2 \left(   \hat{D}_{1,1,2,2} + \hat{D}_{1,2,1,2}  + \hat{D}_{2,1,1,2} \right) \, ,
\end{equation}
where the rational piece $\mathcal{G}^{(0)}_{1,1}$ was fixed in~\cite{Giusto:2018ovt} by checking that the contributions of the identity and the $J_3$ current in the $s$-channel are correctly normalised: the result can be read from~\eqref{eq:G0k1} below with $k=1$. For general $k$, the $s$-channel result in coordinate space was obtained from the supergravity calculation in the previous section and given in~\eqref{eq:stak1}
\begin{equation}\label{eq:l1corr-s}
\mathcal{G}^{(s)}_{k,1} = 1-\frac{k}{N}+\frac{2 k}{N\,\pi}\frac{|1-\alpha_c z|^2}{|1-\alpha_c|^2}|1-z|^2 \left( \sum_{p=1}^k  {1 \over p} \hat{D}_{p,p,2,2} \right) \, . 
\end{equation}
A consistent proposal for the complete correlator, including the $t$ and $u$-channel contributions, is 
\begin{equation}\label{eq:l1corr}
\mathcal{G}_{k,1} =  \mathcal{G}^{(0)}_{k,1}+ \frac{2 k}{N\,\pi}\frac{|1-\alpha_c z|^2}{|1-\alpha_c|^2}|1-z|^2 \left( \sum_{p=1}^k  {1 \over p} \hat{D}_{p,p,2,2} + \hat{D}_{k,k+1,1,2}  + \hat{D}_{k+1,k,1,2} \right) \,.
\end{equation}
The rational contribution $\mathcal{G}^{(0)}$ depends on the precise definition of the external states: a natural choice is to work with supergravity fields that have vanishing extremal couplings, which corresponds, on the CFT side, to redefine the dual operators with double-trace corrections suppressed by $1/\sqrt{c}$ (see for instance~\cite{Arutyunov:2000ima,Rastelli:2017udc}); in this case we have
\begin{equation}
  \label{eq:G0k1}
   \mathcal{G}^{(0)}_{k,1} = \left( 1-\frac{k}{N}\right) \left(1+\frac{\sigma}{k} |1-z|^2 +\frac{\tau}{k} \frac{|1-z|^2}{|z|^2}\right)\,.
\end{equation}
Of course there are other possible $k$-dependent generalizations that reduce to (\ref{eq:11corr}) for $k=1$, but~\eqref{eq:l1corr} is the only one that reproduces correctly the expected result in the following OPE limit: consider the case $\alpha_c \to -\infty$, then the leading contribution to the $z\to 0$ and $z\to \infty$  limit is due to the exchange of a BPS double-trace operator and so has to match the result obtained in the free orbifold description. The free CFT result can be derived for instance from (A.13)-(A.14) of~\cite{Bombini:2017sge}
\begin{equation}
\mathcal{G}_{k,\mathrm{free}}(-\infty,-\infty,z,\bar z)=1+\frac{1}{N}\left[\frac{|z|^{\frac{2}{k}}-|z|^2}{1-|z|^{\frac{2}{k}}}+\frac{1+|z|^2+|1-z|^2}{2}-k\right]\,.
\end{equation}
In the $z\to 0$ limit one has
\begin{equation}
\mathcal{G}_{k,\mathrm{free}}(-\infty,-\infty,z,\bar z)\to 1+\frac{1-k}{N} ~.
\end{equation}
In order to reproduce the same result from~\eqref{eq:l1corr}, we need to fix the coefficient of $\hat{D}_{k+1,k,1,2}$ and $\hat{D}_{k,k+1,1,2}$ as done above. This result also yields a very compact answer in Mellin space:
\begin{equation}
  \label{eq:m1corr}
  \widetilde{\cal M}_{k,1} = -{2 k \over  \Gamma(k) } \left( {1 \over s} + {1 \over t - k+1} +  {1 \over \tilde{u} - k+1}  \right)\;.
\end{equation}
Notice that the poles of $\widetilde{\cal M}_{k1}$ are the ones expected from the CFT OPE analysis. To perform this check, one needs to reconstruct the Mellin transform ${\cal M}$ of the full correlator $\mathcal{G}$ from that of the stripped amplitude given in~\eqref{eq:m1corr}. One then finds that ${\cal M}$ has a pole for either $t$ or $u$ equal to $k-1$, depending on the choice of the R-symmetry variables, which corresponds to the exchange of single-trace CPOs of twist $k-1$ that can appear in the OPE between the two single-trace CPOs $O^{(k)}$ and $O^{(1)}$. Since these higher twist operators are absent in the OPE of $O^{(1)}$ with $O^{(1)}$, one expects only the pole at $s=0$; this is indeed what happens in~\eqref{eq:m1corr}, thanks to the fact that all other $s$-channel poles non-trivially cancel out after performing the sum over $p$ in~\eqref{eq:l1corr}.

As for general $l$, we take some inspiration from the Mellin amplitudes of ${\rm AdS}_5 \times S^5$, as well as the structures of the flat-space amplitudes of ten and six-dimensional supergravity theories. The tree level Mellin amplitudes of ${\rm AdS}_5 \times S^5$ were determined in~\cite{Rastelli:2016nze}  and take the following form\footnote{For the most general case, $\langle O_{p_1} {O}_{p_2} O_{p_3} {O}_{p_4} \rangle$, the Mellin amplitude is given by a similar form, here we concern ourselves with $\widetilde{\cal M}_{kl}$.} (see~(25) of~\cite{Rastelli:2016nze}) 
 \begin{align}   \label{eq:ads5-corr}
 \widetilde{\cal M}^{\rm AdS_5}_{k,l}(s,t; \sigma, \tau)   = \sum_{p+q+r=l-1} \left(  {  a_{pqr} \,  \sigma^q \tau^r  \over (s-s_M+2p)(t-t_M+2r)(\tilde{u}-{u}_M+2q) } \right) \, ,
 \end{align}
 with $s_M = {\rm min} \{p_1+p_2, p_3 +p_4\}-2=2(l-1)$, $t_M  = {\rm min} \{p_2+p_3, p_1 +p_4\}-2=k{+}l{-}2$, and ${u}_M = {\rm min} \{p_1+p_3, p_2 +p_4\}-2=k{+}l{-}2 $. The Mellin amplitudes of ${\rm AdS}_3 \times S^3$ certainly cannot take the same form as \eqref{eq:ads5-corr}. In particular, the flat-space limits of the Mellin amplitudes in ${\rm AdS}_5 \times S^5$ and the Mellin amplitudes in ${\rm AdS}_3 \times S^3$ should have different structures. Indeed, the four-point amplitude of tensor multiplets in six-dimensional $(2,0)$ supergravity (see e.g.~(27) of~\cite{Heydeman:2018dje}) and the four-point amplitude of ten-dimensional type IIB supergravity (see e.g.~(1.1) of~\cite{Boels:2012ie}) take the following distinguished forms: 
 \begin{align} \label{eq:flat-space}
 A_4^{(2,0)}  = \delta^{8}(Q) \left( {\delta_{f_1 f_2}\delta_{f_3 f_4} \over s} + {\delta_{f_2 f_3}\delta_{f_1 f_4} \over t} + { \delta_{f_1 f_3}\delta_{f_2 f_4} \over u} \right) \, , \qquad A_4^{\rm IIB} = {\delta^{16}(Q) \over s t u}   \, ,
 \end{align}
 where $f_i$ are the flavour indices, and $s, t, u$ are the standard Mandelstam variables. Here $\delta^{8}(Q)$ and $\delta^{16}(Q)$ are for supersymmetry, their precise forms are not important for our discussion. The amplitudes with the overall $\delta^{8}(Q)$ and $\delta^{16}(Q)$ removed are the analogue of the stripped Mellin amplitudes $\widetilde{\mathcal{M}}$. 
 
Some comments are in order. In the type IIB theory compactified on $T^4$, the four-point tree-level amplitude of 6D $(2,2)$ supergravity  takes the form 
 \begin{align}
 A_4^{(2,2)} = {\delta^8(Q)\delta^8(\tilde Q) \over stu} \,,
 \end{align}
 which is very much similar  to the 10D amplitude $A_4^{\rm IIB}$ given in above. However, due to the maximal supersymmetry, the corresponding on-shell superfield contains both the graviton multiplet and the tensor multiplet, whereas the operators of the correlation functions we consider are dual to the tensor multiple only. To extract the tensor multiplet part, one can perform Grassmann variable integration (as shown in \cite{Heydeman:2018dje}), which effectively replaces $\delta^8(\tilde Q)$ by $-{1\over 2}(s^2 + t^2 +u^2)$, and leads to $ \delta^{8}(Q) \left( {1 \over s} + {1 \over t} + { 1 \over u} \right) $. Therefore, the component amplitude of tensor multiplets of $A_4^{(2,2)}$ (relevant for the correlation functions we consider here) is identical to $ A_4^{(2,0)}$ when all the flavours $f_i$ are the same.\footnote{We certainly expect to see the difference at loop level, since the states running in loops will be different for $(2,2)$ and $(2,0)$ supergravity, which however cannot happen for tree diagrams.}  
 
Comparing the flat-space amplitudes in~\eqref{eq:flat-space} with the ${\rm AdS}_5 \times S^5$ Mellin amplitudes in~\eqref{eq:ads5-corr} (as well as the results of the special cases of $k=l$ in \eqref{eq:kk-stu-2}) motivates us to propose the Mellin amplitudes in ${\rm AdS}_3 \times S^3$ to take following general form\footnote{The coefficients $a_{pqr}$ in (\ref{eq:ads5-corr}) for the case of ${\rm AdS}_5\times S^5$ were determined in~\cite{Rastelli:2016nze}, up to overall normalizations, by requiring that the residue of ${\cal M}$ at each pole be a polynomial due to locality. The same procedure unfortunately cannot apply here, since, by construction, this condition does not impose any constraints on $b_{pqr}$ in (\ref{eq:mklcorr}).}
        \begin{align}   \label{eq:mklcorr}
 \widetilde{\cal M}_{k,l}(s,t; \sigma, \tau) = \sum_{p+q+r=l-1} b_{pqr} \,  \sigma^q \tau^r  \left(  { \delta_{f_1 f_2}\delta_{f_3 f_4}  \over s-s_M+2p} +{ \delta_{f_2 f_3}\delta_{f_1 f_4} \over t - t_M + 2r } +{ \delta_{f_1 f_3}\delta_{f_2 f_4} \over \tilde{u} - {u}_M+ 2q }  \right) \, ,
             \end{align}   
where $s_M, t_M$ and $u_M$ are the same as those appearing in the case of ${\rm AdS}_5 \times S^5$ as defined in~\eqref{eq:ads5-corr}.              
The correlator $\widetilde{\cal M}_{k,l}$ is symmetric under $1 \leftrightarrow 2$, which constrains $b_{pqr}  = b_{prq}$. From the above general ansatz and the $s$-channel result given in \eqref{eq:Ms-kl}, we can uniquely fix the coefficient $b_{pqr}$, which leads to     
  \begin{align} \label{eq:apqr}
b_{pqr} = - {2 k l \over \Gamma(k)\Gamma(l)} {l-1\choose p,q,r}^2 \,  \left( \prod^{q+r}_{i=1} { k{-} i \over l {-} i  } \right)  \, ,
 \end{align}       
where, as in \eqref{eq:Ms-kl}, $\prod^{0}_{i=1} \left(\ldots\right) \equiv 1$.
For $l=1$ we see that \eqref{eq:mklcorr}-\eqref{eq:apqr} reproduces the result given in~\eqref{eq:m1corr}, and for $k=l$ it reduces to \eqref{eq:kk-stu-2}. Furthermore, $b_{pqr}$ given in \eqref{eq:apqr} manifestly satisfies the constraints $b_{pqr} = b_{prq}$, which may be viewed as another consistency check.  

\section{Conclusions}
\label{sec:conclusions}

In this paper we focused on the AdS$_3$/CFT$_2$ duality relevant to type IIB string theory on AdS$_3\times S^3\times \mathcal{M}$, where $\mathcal{M}$ is either $T^4$ or $K3$. We considered the four point correlators~\eqref{eq:LLLLcor} between CPO operators of left/right dimension $(k/2,k/2)$, with $k=1,2,\ldots$, which are dual to supergravity fields in the $h^{1,1}(\mathcal{M})+1$ tensor multiplets of the theory. As done in~\eqref{eq:Aalpha}, it is convenient to encode the R-symmetry dependence of each CPO in terms of continuous variables $A_\alpha$, so the correlator is determined by a function ${\cal G}_{k,l}$ of the cross-ratios $z$, $\bar{z}$ of the spacetime positions and the cross-ratios $\alpha$, $\bar{\alpha}$ of the R-symmetry variables, see~\eqref{eq:LLLLcor}. As a consequence of the  superconformal Ward identity, ${\cal G}_{k,l}$ satisfies the constraint~\eqref{eq:superWI} and can be split in a simple rational contribution $\mathcal{G}^{(0)}_{k,l}$ and a dynamical $\widetilde{\mathcal{G}}_{k,l}$, as written in~\eqref{eq:theform}. 

In particular in this paper we studied the supergravity correlator with two operators of dimension $k$ and two operators of dimension $l$ and presented explicit formulae for $\widetilde{\mathcal{G}}_{k,l}$. The approach we took is schematically as follows: we started from a generalisation of the results discussed in \cite{Bombini:2017sge} for the Heavy-Heavy-Light-Light correlators and assumed that the light limit captures the $s$-channel of the correlator involving all single trace operators. As summarised at the end of section~\ref{sec:HHLL}, the result is
\begin{equation}\label{eq:2.28bis}
  \widetilde{\mathcal G}^{(s)}_{k,l}(\sigma ,\tau ,z,\bar z) =\frac{2k l^2}{\pi} \sigma^{l-1}\,|1-z|^{2l}\sum_{p=1}^k c_{p,l}(\sigma,\tau) \hat D_{p,p,l+1,l+1}\,,
\end{equation}
where the coefficients $\hat{c}$ are defined in~\eqref{eq:defcp} and $\hat D$-functions are defined in~\eqref{eq:hatDM}.

As it is by now well established, the Mellin space formulation of the holographic correlators (see the definition~\eqref{eq:MTstramp}) takes a simple form, see~\eqref{eq:Ms-kl}. When all the CPOs in the correlator have the same flavour the $s$-channel result discussed above is not the complete answer. In section~\eqref{sec:corr-mellin}, we use the simplicity of the Mellin space formulation to present a formula for the dynamical contribution of the correlators valid also in this case: we first start from the case where all CPOs have the same conformal dimension and use permutation symmetry between the operators to obtain the complete correlator~\eqref{eq:kk-stu-2}; then we move on to the case involving two operators of dimension $k$ and two of dimension $l$ with $k\not= l$. At the end of section~\ref{sec:corr-mellin} we provide an explicit conjecture for this type of correlators, which for convenience we rewrite below
\begin{align}   \label{eq:mklcorrbis}
 \widetilde{\cal M}_{k,l}(s,t; \sigma, \tau) = \sum_{p+q+r=l-1} b_{pqr} \,  \sigma^q \tau^r  \left(  { \delta_{f_1 f_2}\delta_{f_3 f_4}  \over s-s_M+2p} +{ \delta_{f_2 f_3}\delta_{f_1 f_4} \over t - t_M + 2r } +{ \delta_{f_1 f_3}\delta_{f_2 f_4} \over \tilde{u} - {u}_M+ 2q }  \right) \, ,
             \end{align}   
where in our case $s_M = 2(l-1)$, $t_M = {u}_M = k + l - 2$, $f_i$ are the flavour indices of the four CPOs and 
  \begin{align} \label{eq:apqrbis}
b_{pqr} = - {2 k l \over \Gamma(k)\Gamma(l)} {l-1\choose p,q,r}^2 \,  \left( \prod^{q+r}_{i=1} { k{-} i \over l {-} i  } \right)  \, ,
 \end{align}       
 with $\prod^{0}_{i=1} \left(\ldots\right) \equiv 1$. This expression has the expected structure in Mellin space and satisfies the simple OPE checks we performed. Of course it would be interesting to provide further checks of~\eqref{eq:mklcorrbis}~. One approach is to use the same supergravity technique adopted here, but starting from a more general geometry that, in the light limit, produces the insertions $O^{(k_1,f)}_1 O^{(k_2,f)}_2$. Then the results obtained in this way can be used to justify the form of the correlators $\langle O^{(k_1,f)}_1 O^{(k_2,f)}_2 O^{(k_3,g)}_3 O^{(k_4,g)}_4 \rangle$ in all channels. We plan to discuss the details of the supergravity derivation of the general correlators in a separate work~\cite{GRTW}. In~\cite{Rastelli:2019gtj} these AdS$_3$ correlators were studied by taking a different approach where both Witten diagrams and a hidden conformal symmetry were used: in all explicit checks we performed, we found that the our results for $\widetilde{\mathcal G}^{(s)}_{k,l}$ and those in~\cite{Rastelli:2019gtj} agree. This strongly supports our assumptions on the smoothness of the light limit of the HHLL correlators and on the $t$ and $u$-channel contributions for the correlators involving operators with the same flavour.

%%%%%%%%%%%%%%%%%%%%%%%%%%%%%%%%%%%%%%%%%%%%%%%%%%%%%%%%%%%%
\section*{Acknowledgements}

We would like to thank Alessandro Bombini, Nej\v{c} Ceplak, and Andrea Galliani for discussions and collaboration on a related project. This work was supported in part by the Science and Technology Facilities Council (STFC) Consolidated Grant ST/P000754/1 {\it String theory, gauge theory \& duality} and by the MIUR-PRIN contract 2017CC72MK003. C.W. is supported by a Royal Society University Research Fellowship No. UF160350.

%\bibliographystyle{utphys}      % (uses file "utphys.bst")
%\bibliography{microstates2}              % expects file "microstates.bib"

\providecommand{\href}[2]{#2}\begingroup\raggedright\endgroup

\end{document}